# THERMOPHYSICAL CHARACTERISTICS OF THE POROUS SILICON SAMPLES FORMED BY ELECTROCHEMICAL, CHEMICAL AND COMBINED ETCHING METHODS

B. Zhumabay[1*], R. Dagarbek[2], D. Kantarbayeva[1], I. Nevmerzhitsky[1], B. Rakymetov,[1] A. Serikkanov[1], Zh. Alsar[1], K.B. Tynyshtykbayev[1,3], and Z. Insepov[4*]

[1] Satbayev University, Physics and Technology Institute, Almaty 050032, Kazakhstan
[2] Almaty University of Energy and Communications, Almaty 050032, Kazakhstan
[3] Nazarbayev University, Astana 010000, Kazakhstan
[4] Purdue University, West Lafayette, Indiana, United States of America

* bzhumabay05@gmail.com
* zinsepov@purdue.edu

**Abstract.** The article addresses the thermophysical properties of porous silicon (PS) samples produced through electrochemical (EC), metal-assisted chemical (MACE), and combined (MACE + EC) etching methods. The PS/MACE+EC sample's thermophysical properties exhibited higher values than both the PS/EC and PS/MACE. The energy activity associated with the process of water splitting on the surface of porous silicon (PS) formed by the metal-assisted chemical etching (MACE) combined with the electrochemical etching (EC) method is found to be more significant compared to the PS samples prepared using only EC or MACE techniques. The combined MACE/EC method is considered the most effective technique for obtaining PS with a nanoporous silicon surface, which has the highest energy activity in water-splitting processes. The activation energy required for water splitting, denoted as Ea (PS/MACE+EC), exhibits a higher value compared to the activation energies observed for the Ea (PS/EC) and Ea (PS/MACE) samples. The inclusion of nickel (Ni) particles within the pores of porous silicon (PS<Ni>) serves to stabilize the thermophysical properties of the material. Consequently, this prevents substantial alterations in the thermal characteristics of PS<Ni> compared to PS without nickel in its pores. The samples containing nickel (PS/MACE) and PS/MACE+EC exhibit long-term preservation of their energy activity (Ea) for one year. This phenomenon occurs due to the presence of nickel-containing nanopores within the samples.
The article additionally examines the thermophysical properties of porous silicon (PS) in comparison to crystalline silicon (c-Si) and powdered silicon (powder-Si).

## Introduction

As it is well known, the porous silicon (PS) has a high catalytic activity of hydrogen evolution during the photoelectrochemical splitting of water [1]. For example, using a surface-enriched Ni layer on the surface of a porous silicon photocathode PS<Ni> made it possible to increase its catalytic activity two times higher than the catalytic activity of the Pt- electrode in the reaction of hydrogen evolution [2].

This work compares the thermophysical characteristics of PS samples formed by EC [3], MACE [4], and combined (MACE + EC) etching methods with the thermal characteristics of crystal silicon (c-Si) and powder silicon (powder-Si) samples.

**Materials and methods.** The wafers of single-crystal silicon c-Si<100>, p-type (boron), with an ohm of 10 Ωcm and thickness of 500 μm, were used in the experiment. To form an ohmic contact, the Al layer (300 nm) was deposited on the Si (100) wafer's rear polished surface by magnetron sputtering, followed by annealing at T=500°C for 30 min in an inert Ar medium. The PS formation was reached using the following methods, Table 1.

Table 1. Electrolyte composition in various etching methods

| Method # | Electrolyte abbreviation | Electrolyte composition | Method (sample) abbreviation |
|---|---|---|---|
| I | Et1 | $HF:H_2O_2=1:1$ | EC |
| II | Et2 | $HF:H_2O_2: Ni(NO_3)_2 =1:1:1$ | MACE |
| III | Et3 | $HF:H_2O_2: Ni(NO_3)_2 =1:1:1$ | EC + MACE |

At the MACE method, the maximum formation of pores occurs in the first 30 seconds. The further increase of the etching time to 30 minutes does not affect the formation of pores. Etching was carried out in fluoroplastic cells.

The following experimental techniques: the SEM Carl Zeiss Auriga Crossbeam 540 with an X-ray energy-dispersion analyzer; Raman spectrometer, the Horiba LabRam Evolution; the STA6000/8000 calorimeter of differential scanning calorimetry (DSC); and thermogravimetric analysis (TGA) were used. The dynamics of water splitting on the surface of PS samples at room temperature was studied by weighing the samples' mass using a ScoutPro 2000g/0.1 g electronic balances.

**Results and discussion.**

**Scanning electron microscopy.** SEM was applied to study the influence of the etching method on the surface morphology of porous silicon (PS) samples. Results show that the combined method (MACE+EC) gives the highest porosity of PS samples. At the same time, the metal-assisted chemical etching (MACE) method gives the smallest percentage of pores on PS samples. Figure 1 shows SEM- images of samples: PS/EC – a), PS /MACE – b), PS/MACE + EC – c).

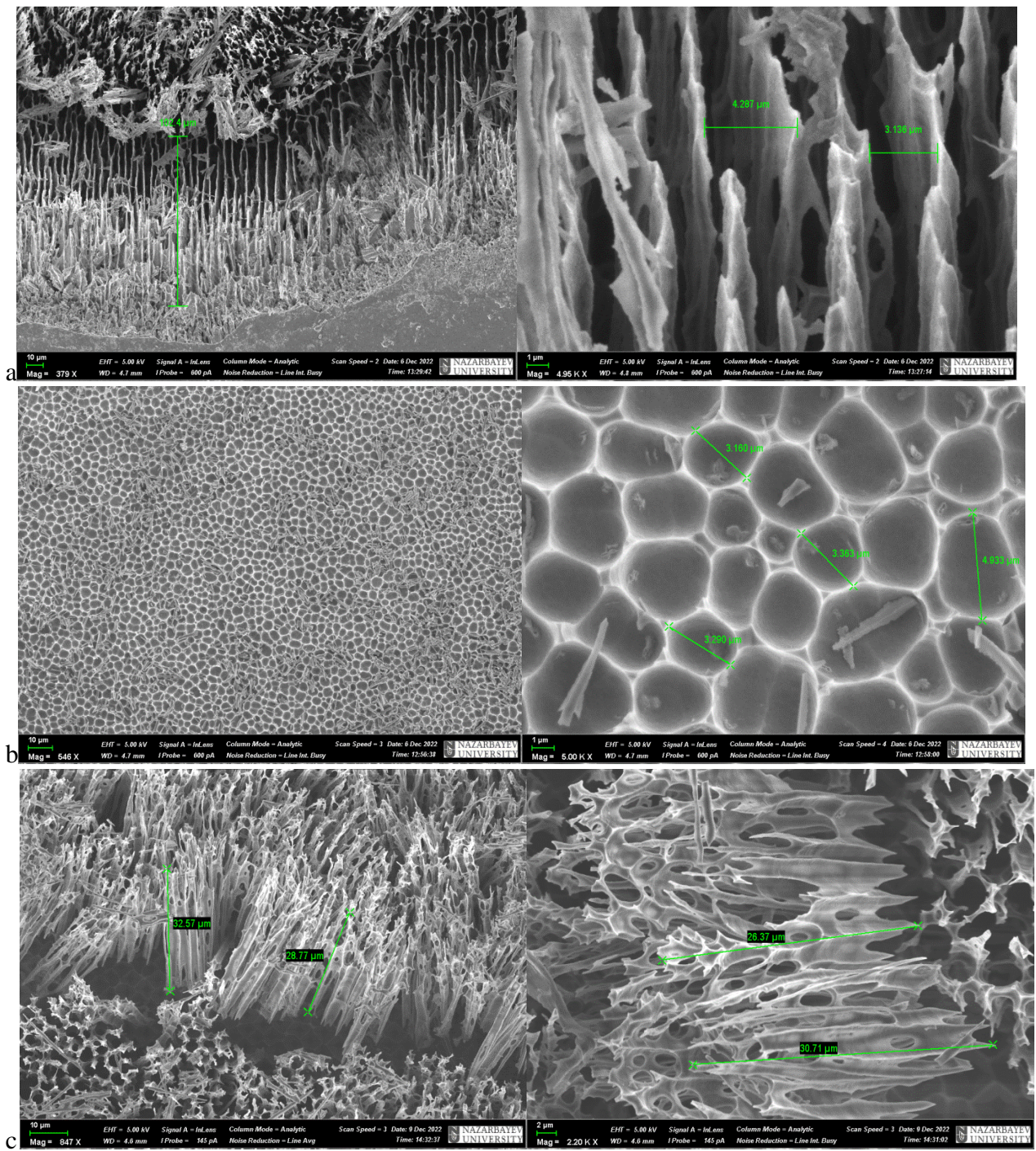

**Figure 1. SEM- images of samples PS/EC –a), PS/MACE –b), PS/MACE+EC –c).**

**The highest porosity is observed on the PS/MACE+EC samples, Figure 1c, the smallest - on the PS/MACE sample formed by the MACE- method, Figure 1b.**

**Raman spectra.** The Appendix of this article presents the Raman spectra images of the samples being investigated. The Figure 1A (Appendix) shows that the samples with higher porosity have a higher intensity of the Raman peak at a high-frequency shift. The highest intensity of the Raman peak corresponds to $I_v$ = 54,000 count at a frequency of $v_{PS/MACE+EC}$ = 520.32 cm$^{-1}$ and it is observed for the PS/MACE+EC sample formed by the (MACE + EC) combined etching method in the Et3 electrolyte (Figure 1A - d). For the PS/MACE (Et2) sample the following values of intensity and frequency are observed: $I_v$ = 33,000 count at $v_{PS/MACE}$ = 519.66 cm$^{-1}$ (Figure 1A - c). The sample PS/EC (Et1) is characterized by $I_v$ = 32,000 count, $v_{PS/EC}$ = 519.11 cm$^{-1}$ (Figure 1A - b). The Raman peak intensity for the crystalline c-Si is $I_v$ = 7000 count at $v_{c-Si}$ = 519.11 cm$^{-1}$ (Figure 1A -a).

The Raman spectra of crystalline (c-Si) and porous (PS) silicone have a characteristic resonance peak due to electron-phonon scattering, the intensity and frequency of which depend on the concentration of scattering centers and, consequently, on the conditions for the formation of pores. The high-frequency shift and decrease in the full width at half maximum (FWHM) of the main peak, (increase in peak intensity) for PS/MACE+EC up to the frequency $v_{PS/MACE+EC}$ = 520.32 cm$^{-1}$ lead to an increase in the number of Raman scattering centers in the form of nc-Si nanocrystallites [5].

**The energy dispersive X-ray analysis.** The EDS-spectra of samples are shown in Appendix (Figure 2A). The characteristic X-ray Ni lines in the EDS-spectrum of the PS/MACE sample appear in the presence of Ni-lines 0.2 Wt% (Figure 2A – c). The PS/MACE+EC sample has the Ni lines weakly pronounced at the background level (0.1 Wt%), Figure 2A – b. In PS/EC samples, the characteristic X-ray Ni- lines in the EDS- spectrum are not observed (Figure 2 – a). The MACE-method gives the most significant presence of Ni in pores.

**Calorimetry.** The calorimetric characteristics of initial c-Si, then PS/EC, PS/MACE, PS/MACE+EC, and powder-Si samples were analyzed using enhanced TGA and DSC techniques. The investigation focused on the changes in calorimetric features under isobaric temperature changing (ΔT) conditions, with a temperature range from Tonset = 30°C to Tend = 900°C and a heating rate of 10°C/min.  The measurements were carried out in nitrogen gas and compressed air environments. Gas consumption - 20.0 ml/min. The TGA and DSC spectra are given in the Appendix (Figures 3A-7A).

The characteristic TGA and DSC spectra in an N$_2$-atmosphere for the PC/EC sample are shown on the Figure 2.

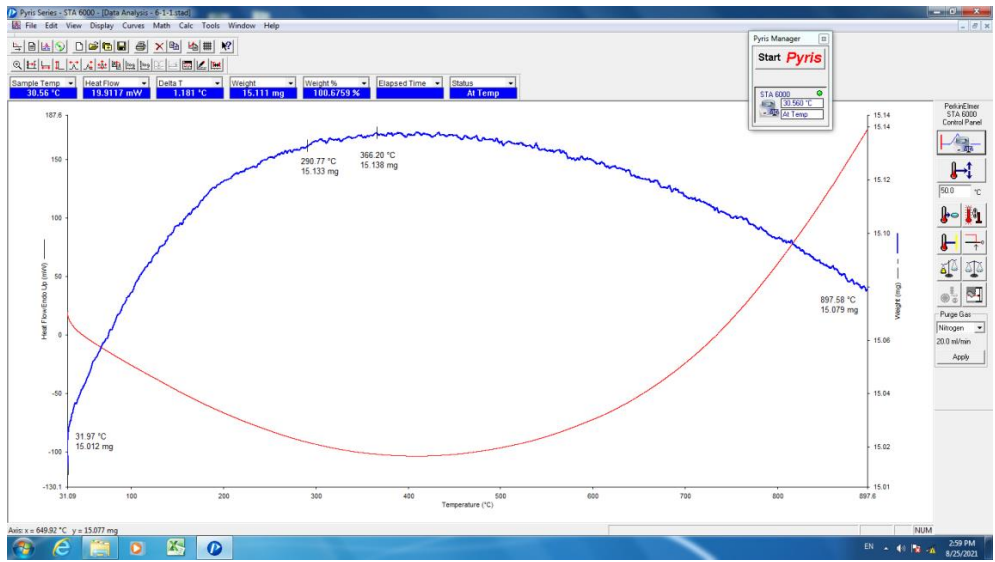

Figure 2. TGA (blue line) and DSC (red) spectra for a PS/EC sample in an $N_2$- atmosphere.

The TGA and DSC spectra record changes in the mass $\Delta m$ (right Y-axis) of the sample and heat flux $\Delta Q$ (left Y-axis) with an isobaric temperature change $\Delta T$ (X-axis) at the points of characteristic temperatures $T_{onset}$, $T_{peak}$, and $T_{end}$. Two typical regions appear on the TGA and DSC spectra: I – an increase of the mass sample $\Delta m_{1-2}$ (on the TGA curve) with a decrease in the heat flux $\Delta Q_{1-2}$ (on the DSC curve); II - a reduction in sample weight $\Delta m_{2-3}$ with an increase in heat flux $\Delta Q_{2-3}$. Changes in $\Delta m$ and $\Delta Q$ indicate the occurrence of phase transformations in the sample associated with heat absorption, an endothermic process, I-region on the DSC curve between $T_{onset} = 30°C$ and $T_{peak} = 400°C$, and with heat release, exothermic - II region from $T_{peak} = 400°C$ to $T_{end} = 900°C$. The phase transition temperature is determined through the initial temperature $T_{onset}$, at which the phase transition begins under the assumption that the heat of the process is zero and is described in the zero approximation when there are no chemical reactions and phase transitions [6]. The area $\Delta S$ bounded by the experimental DSC curve and the baseline connecting the points for the characteristic temperatures $T_{onset}$ and $T_{end}$ on the DSC curve is proportional to the heat of the entire reaction $\Delta Q_{1-3}$. The aforementioned conditions are satisfied in processes that occur without extraction from the measuring cell, which applies to our specific situation.

The heat flux is defined as

$$\Phi = \Delta Q/\Delta t = C_p \Delta T /\Delta t, \qquad (1)$$

where $\Delta t = t_{start} - t_{finish}$ is the reaction time between the start $t_s$ and the finish $t_f$ of the corresponding reactions in characteristical I- and II- regions of the TGA and DSC curves, respectively. $\Delta T$ is the difference between the corresponding temperatures $\Delta T(I) = T_{onset} - T_{peak}$ and $\Delta T(II) = T_{peak} - T_{end}$.

The heat capacity is

$$C_p \,(J/gK) = \Delta Q \Delta t / \Delta m \Delta T. \qquad (2)$$

The values of changes in the mass Δm, heat flux ΔQ, as well as the meanings of the heat capacities $C_p(I)$, $C_p(II)$ according to (2) of the samples are given in Table 2.

Table 2. Calorimetric parameters of c-Si, powder-Si, PS/EC, PS/MACE, and PS/MACE+EC samples in nitrogen and compressed air.

| Sample | Nitrogen | | | | Air | | | |
|---|---|---|---|---|---|---|---|---|
| | $\Delta m_{1-2}$, (I) $\Delta m_{2-3}$, (II) $\Delta m_{1-3}$, (III) mkg | $\Delta Q_{1-2}$, (I) $\Delta Q_{2-3}$, (II) $\Delta Q_{1-3}$, (III) mW | $T_{peak}$, °C | $C_p$, KJ/gK | $\Delta m_{1-2}$, (I) $\Delta m_{2-3}$, (II) $\Delta m_{1-3}$, (III) mkg | $\Delta Q_{1-2}$, (I) $\Delta Q_{2-3}$, (II) $\Delta Q_{1-3}$, (III) mW | $T_{peak}$, °C | $C_p$, KJ/gK |
| c-Si | (+) 40 (-) 176 (-) 136 | (-) 138 (+) 183 (+) 45 | 410 | 20.7 (I) 6.24 (II) 1.98 (III) | (+) 74 (-) 91 (-) 17 | - - (-) 182 | 910 | - (I) - (II) 64.23(III) |
| powder-Si | (+) 65 (-) 109 (-) 44 | (-) 142 (+) 229 (+) 87 | 420 | 13.11 (I) 12.60 (II) 11.86 (III) | (+) 116 (-) 43 (+) 73 | - - (-) 240 | 910 | - - 19.73 (III) |
| PS/AE | (+) 128 (-) 61 (+) 67 | (-) 120 (+) 287 (+) 167 | 400 | 5.62 (I) 28.23 (II) 14.95 (III) | (+) 74 (-) 117 (-) 43 | (-) 135 (+) 178 (+) 43 | 420 | 10.95 (I) 9.13 (II) 6.00 (III) |
| PS/ MACE | (+) 61 (-) 113 (-) 52 | (-) 130 (+) 320 (+) 190 | 380 | 12.79 (I) 16.99 (II) 21.92 (III) | (+) 70 (-) 125 (-) 55 | (-) 160 (+) 70 (-) 90 | 475 | 13.71(I) 3.36 (II) 9.82 (III) |
| PS/ MACE +AE | + 76 - 110 -34 | (-) 140 (+) 100 (-) 40 | 450 | 11.05 (I) 5.45 (II) 7.06 (III) | (+) 120 (-) 71 (+) 49 | (-) 257 (+) 7 (-) 250 | 750 | - - 31.47 |

An analysis of the calorimetric characteristics (Table 2) and the TGA and DSC spectra (Figures 3A – 7A, Appendix) shows that c-Si (Figure 3A) and powder-Si (Figure 4A) have almost the same patterns of change in thermal conductivity, in contrast to the spectra of porous silicon, Figures5A - Fig.7A. The PS/EC, PS/MACE samples have the maximum change in heat flux $Q_3$ (PS/EC) = 187.6 mW (Figure 5A), $Q_3$ (PS/MACE) = 231.6 mW (Figure 6A), in contrast to powder-Si $Q_3$ (powder-Si) = 107.1 mW (Figure 4A) and c-Si $\Delta Q_3$ (c-Si) = 65.02 mW, Figure 3A. This dependence, which is responsible for the heat flux ΔQ change in the high-temperature range, is associated with the maximum developed surface PS [7].

Figure 4A demonstrates that the nickel-free porous silicon PS/EC samples exhibit the highest observed changes in sample mass (Δm1-3) and heat flux (ΔQ1-3) when subjected to an isobaric temperature change (ΔT). The stabilization of thermophysical parameters in porous silicon (PS) is attributed to the presence of nickel (Ni) within the pores of PS/MACE and PS/MACE+EC. This prevents notable alterations in the thermal characteristics of PS containing nickel (PS<Ni>), as opposed to PS without nickel in its pores. The initial increase in the sample mass Δm in the I - region is explained by the gas adsorption processes from the medium and the heating of the surface of the samples. Samples of porous silicon have the next porosity: PS/MACE+EC (80%), PS/EC (70%), and PS/MACE (50%). By this, the maximum mass grow $\Delta m_{1-2}$ = 128 μg (0.85%) is observed for the PS/EC sample, then for powder-Si $\Delta m_{1-2}$ = 65 μg (0.277%) and the smallest for c-Si $\Delta m_{1-2}$ = 40 μg (0.149%). In this case, the value $T_{max}$ is changed, at which have place the maximum increment of Δm sample. For PS/EC $T_{max}$ = 400°C (Figure 3A), powder-Si – $T_{max}$ = 250 °C (Figure 2A), c-Si Tmax = 220 °C (Fig. 3A). This is because the saturation of a larger surface occurs at a larger $T_{max}$. With a further increase in the heating temperature, II-region, there is a decrease in the recorded mass of $\Delta m_{2-3}$ samples. For $\Delta m_{2-3}$ (PS/EC) = 61 μg (0.40%); $\Delta m_{2-3}$(powder-Si) =109 μg (0.25%), for $\Delta m_{2-3}$(c-Si) = 176 μg (1.22%). The reduction in the $\Delta m_{2-3}$ of the samples is associated with the processes of removing gases from the samples' surface. The most minor $\Delta m_{2-3}$ decrease is observed for PS/EC, the largest for c-Si, indicating that the pores' gases are more strongly bound than on the surface of c-Si. And at the same time, in porous silicon PS/EC at 900 °C, an increase in the final mass $m_3$ by $\Delta m_{1-3}$ = 67 μg relative to the initial mass $m_1$ = 14945 μg is observed due to the amount of gas adsorbed in the pores.

At 900 °C for the powder-Si and c-Si samples, a decrease in the final mass $m_3$ regarding the initial mass $m_1$ is observed. It is due to additional gas desorption at these high temperatures, for powder-Si $\Delta m_{1-3}$ = (-) 44 μg or by 0.101% to the initial mass $m_1$, and the mass of c-Si decreased by $\Delta m_{1-3}$ = (-) 136 μg or by 0.945% relatively to $m_1$.

An increase in the mass $\Delta m_{1-2}$ in I-region is accompanied by a decrease in the recorded power of the heat flux Q, which indicates the endothermic nature of the process during heat absorption. In II-region, the opposite process is observed, a decrease in the mass $\Delta m_{2-3}$ is accompanied by an increase in the recorded power of the heat flux Q, indicating the process's exothermic nature with heat release. At the same time, the amount of absorbed heat $\Delta Q_{1-2}$(I) in I- region is less than $\Delta Q_{2-3}$(II), the exothermic heat released in II- region, Table 2. This indicates a higher energy activity of the exothermic gas desorption process than the endothermic adsorption process. And the energy of these processes is higher for porous samples PS/EC - $\Delta Q_{1-2}$(I) = (-) 120 mW, $\Delta Q_{2-3}$ (II) = (+) 287 mW, and PS/MAC - $\Delta Q_{1-2}$(I) = (-) 130mW, $\Delta Q_{2-3}$(II) = (+) 320 mW than powder-Si - $\Delta Q_{1-2}$(I) = (−) 142 mW, $\Delta Q_{2-3}$(II) = (+) 229.1 mW and c-Si $\Delta Q_{1-2}$(I) = (−) 138mW, $\Delta Q_{2-3}$(II) = (+) 183.02 mW, where (−) indicates a decrease Q, (+) - an increase Q, Table 2. Energy PS/EC ($Q_3$ = + 187.6 mW, Figure 3A) has 2.77 times higher energy than c-Si ($Q_3 = + 67.807$ mW (Figure 5A). Energy is reduced to the same mass of c-Si and PS/EC samples.

Therefore, it can be observed that the energy of PS/EC is 2.77 times greater than that of c-Si. Significant changes in the character of DSC spectra for crystalline silicon (c-Si) and powder silicon (powder-Si) are observed when the medium's atmosphere is altered to compressed air containing oxygen ($O_2$), as depicted in Figure 3A and Figure 4A, respectively. The observed heat Q exhibits a decrease across the entire range of temperatures, suggesting that the process is endothermic throughout the temperature range of 30–900 °C. Oxidative processes take place in the high-temperature range of 600-800 °C when oxygen is present on the surface of c-Si and powder-Si, resulting in the formation of complexes containing oxygen [8]. The observed change in the DSC spectrum of porous silicon PS/EC in a compressed air atmosphere, as depicted in Figure 5A, exhibits similarities to the DSC spectrum obtained in a nitrogen atmosphere. This similarity suggests the presence of stable SiO2 oxides on the surface of nanocrystallites (nc-Si) in the porous silicon material [5]. The reason for this phenomenon can be attributed to the notable heat capacities of c-Si and powder-Si, which are measured as $Cp(I) = 20.7$ KJ/gK and $Cp(I) = 13.11$ KJ/gK, respectively. Consequently, the ratio of PC/EC, $Cp(I)$ is determined to be 5.62 kJ/gK, as presented in Table 2.

**Water splitting dynamics.** Experiments were conducted to investigate the evaporation of water droplets at room temperature from the surfaces of c-Si, PS/EC, PS/MACE, and PS/MACE+EC samples, as shown in Figure 3.

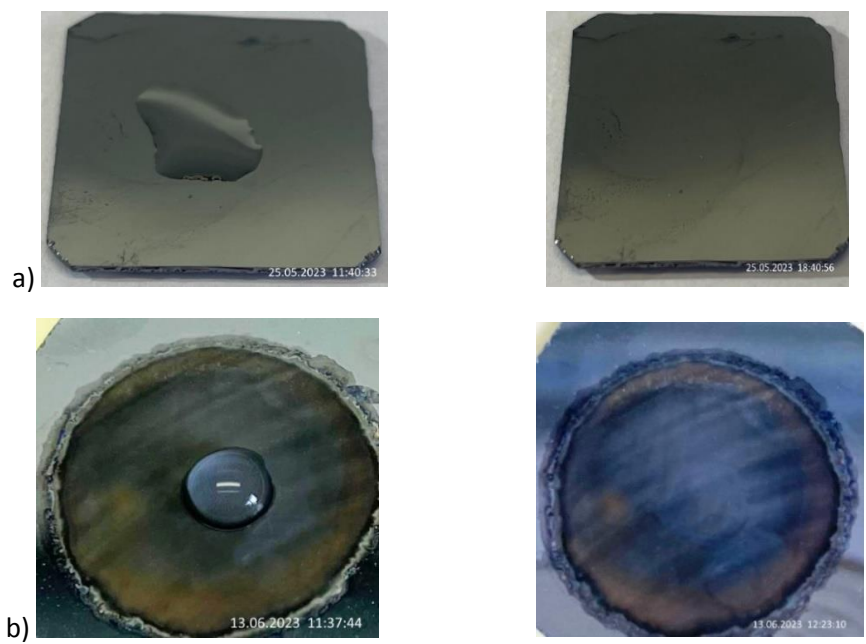

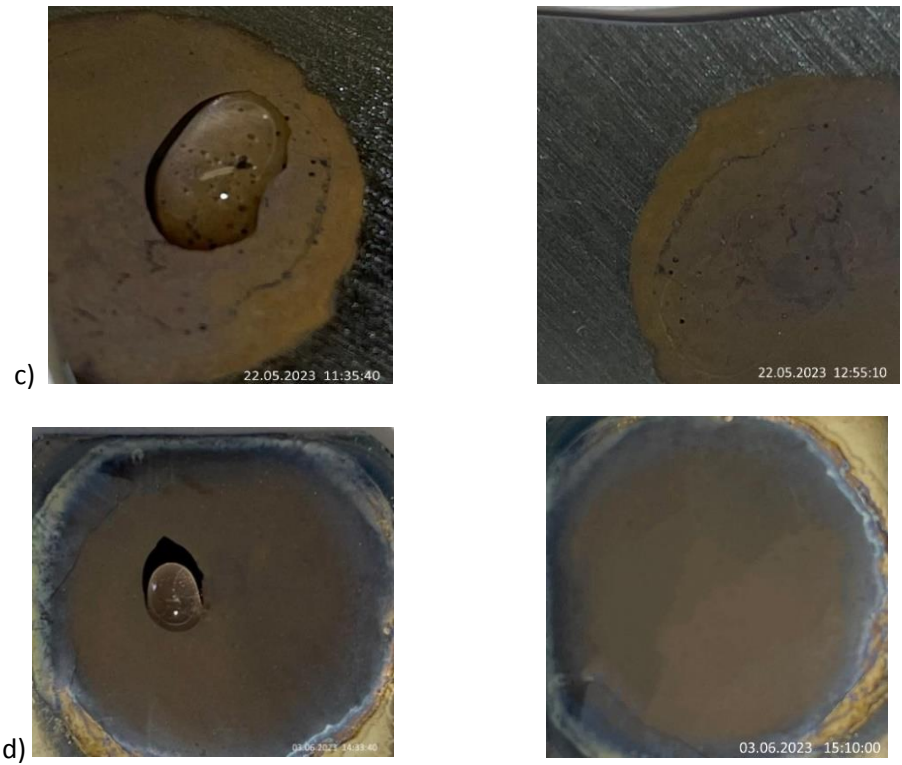

Figure3. Photo of samples: a) c-Si, b) PS/EC, c) PS/MACE, d) PS/MACE+EC. The pictures show the actual time of the experiment, the beginning and end of the evaporation of 1 ml of a water droplet.

The conducted experiments investigated the evaporation of water droplets at room temperature from the surfaces of crystalline and porous silicon samples. These samples were obtained using different etching methods. The results indicated that the PS/MACE+EC samples exhibited the highest energy activity, as depicted in Figure 3d. The evaporation time for a volume of 1 milliliter of water was determined to be $t_{evapor}$ (PS/MACE+EC) = 2180 s. The recorded evaporation times for 1 ml of water were as follows: $t_{evapor}$ (PS/EC) = 2726 s, $t_{evapor}$ (PS/MACE) = 4770 s, $t_{evapor}$ (c-Si) = 25223 s.

The examination of the evaporation dynamics of water molecules, as depicted in Figure 4, reveals that the (PS/MACE+EC) sample exhibits a significantly higher activation energy for the evaporation process ($E_a$ (PS/MACE+EC) = 0.2087 eV) compared to $E_a$(PS/EC) = 0.070 eV, $E_a$(PS/MACE) = 0.050 eV, and $E_a$(c-Si) = 0.025 eV.

The PS/MACE+EC sample exhibited the highest energy activity in comparison to the PS/EC and PS/MACE samples due to its porous surface.

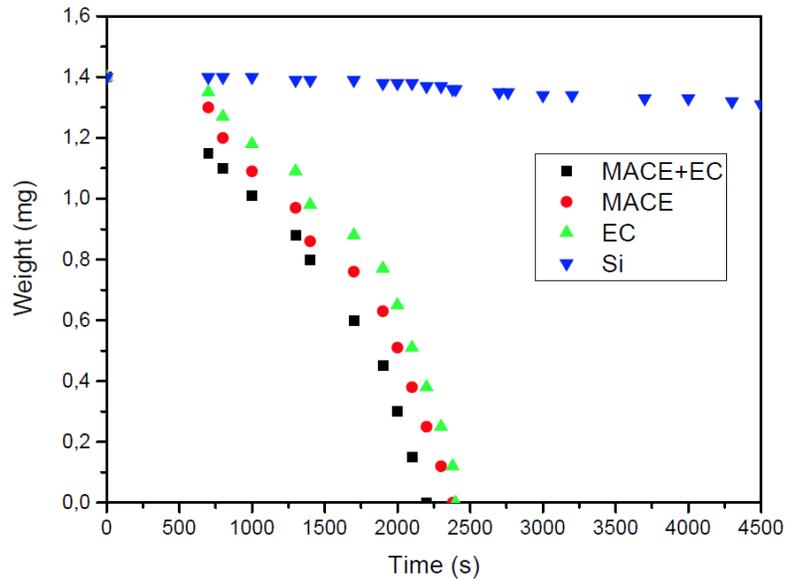

Figure4. The dependence of the water mass on the evaporation time for samples PS/MACE+EC, PS/MACE, PS/EC, c-Si.

The porous surface of the PS/MACE+EC sample has the highest energy activity compared to the PS/EC and PS/MACE samples, which is due to the high porosity of PS/MACE+EC (Figure 1c). In addition, the PS/MACE and PS/MACE+EC samples showed high energy stability over one year in water evaporation experiments.

Figure 5 shows snapshots of freshly etched samples: PS/EC-1, PS/MACE-2, PS/MACE+EC-3 (Figure 5a); and after one year: PS/MACE+EC (Figure 5c) and PS/MACE+EC (Figure 5b) in the experiment on the evaporation of 1 ml water droplet.

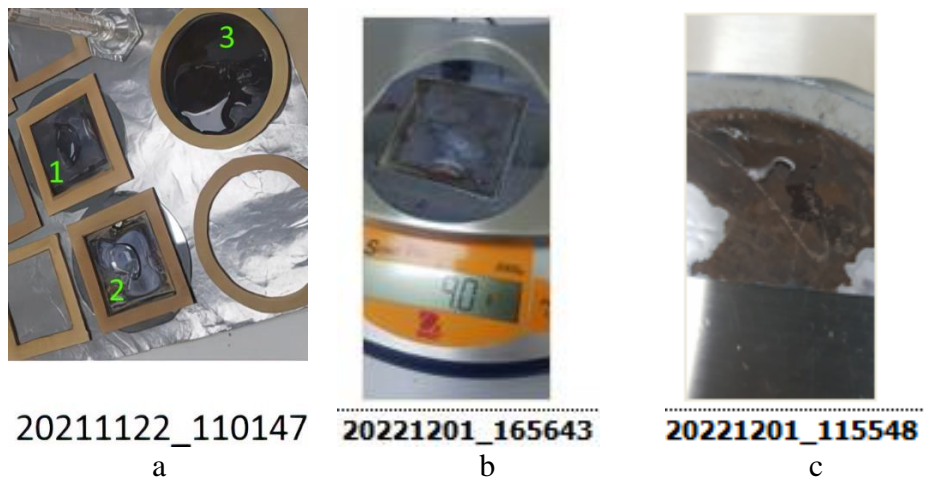

Figure5. Photo a) freshly etched samples: PS/EC-1, PS/MACE-2, PS/MACE+EC-3; b) PS/EC and c) PS/MACE+EC, 1 year later in a 1 ml water droplet evaporation experiment.

The brown color of the PS/MACE+EC sample (Figure 5c) can be attributed to the photoluminescence of nanocrystals of porous silicon (nc-Si PS) [9], indicating its energy activity. In contrast, PS/EC lacks nickel and does not exhibit this characteristic. This finding suggests that the inclusion of nickel in the pores of PS/MACE and PS/MACE+EC enhances the stability of the energy activity of porous silicon. This was observed through calorimetric experiments, which demonstrated that the thermophysical characteristics of nickel-containing porous silicon remain stable, unlike porous silicon without nickel in the pores.

**Main results and conclusion.**

1. The heat capacities of crystalline silicon (c-Si), powdered silicon (powder-Si), and samples of porous silicon (PS) are determined based on the spectra obtained from thermogravimetric analysis and differential scanning calorimeter. These values are established by considering the effects of various etching methods, such as EC-, MACE-, and EC + MACE-etching. The specific heat capacities of $C_p(c\text{-Si})$, $C_p(\text{powder-Si})$, $C_p(\text{PS/MACE})$, and $C_p(\text{PS/AE})$ are 20.7 KJ/gK, 13.11 KJ/gK, 12.79 KJ/gK, and 5.62 KJ/gK, respectively.

2. Applying a hybrid technique involving electrochemical etching combined with Ni-stimulated chemical etching (EC+MACE) enables the production of Ni-containing nanoporous silicon materials that exhibit significant energy activity in water-splitting reactions. The activation energies for the PS/MACE+EC, PS/MACE, and PS/AE systems are 0.2087 eV, 0.070 eV, and 0.050 eV, respectively. The activation energy of water splitting processes, specifically $E_a$ (PS/MACE+EC), is higher than the energy activity of the PS/EC and PS/MACE samples.

3. The inclusion of nickel within the pores of porous silicon (PS/MACE) and PS/MACE+EC leads to the stabilization of their thermophysical parameters. The samples containing nickel (PS/MACE+EC) and $E_a$(PS/MACE) exhibit stability over a period of one year.

**Acknowledgment.** This work was supported by the Ministry of Science and Higher Education of the Republic of Kazakhstan, project BR18574141 "Comprehensive multi-purpose program for improving energy efficiency and resource saving in the energy sector and mechanical engineering for the industry of Kazakhstan" co-funded by Nazarbayev University" SSH2023007, code 064.01.01. The authors express their gratitude to Dr. T.Sh.Atambaev for financial support of the work and to colleagues K.Zhumanova, L. Khamkhash, K.Rustembekova for electron microscope images.

**ORCID**

B. Zhumabay https://orcid.org/0000-0001-8603-6335
R. Dagarbek https://orcid.org/0000-0001-5737-8723
D. Kantarbayeva https://orcid.org/0000-0003-3891-8733
I. Nevmerzhitsky https://orcid.org/0000-0003-4352-8684
B. Rakhmetov https://orcid.org/0000-0002-7326-4774
A. Serikkanov https://orcid.org/0000-0001-6817-9586
Zh. Alsar https://orcid.org/0000-0001-7287-7555
K.B. Tynyshtykbayev https://orcid.org/0000-0002-4761-7409
Z. Insepov https://orcid.org/0000-0002-8079-6293

**Appendix**

The Raman spectra

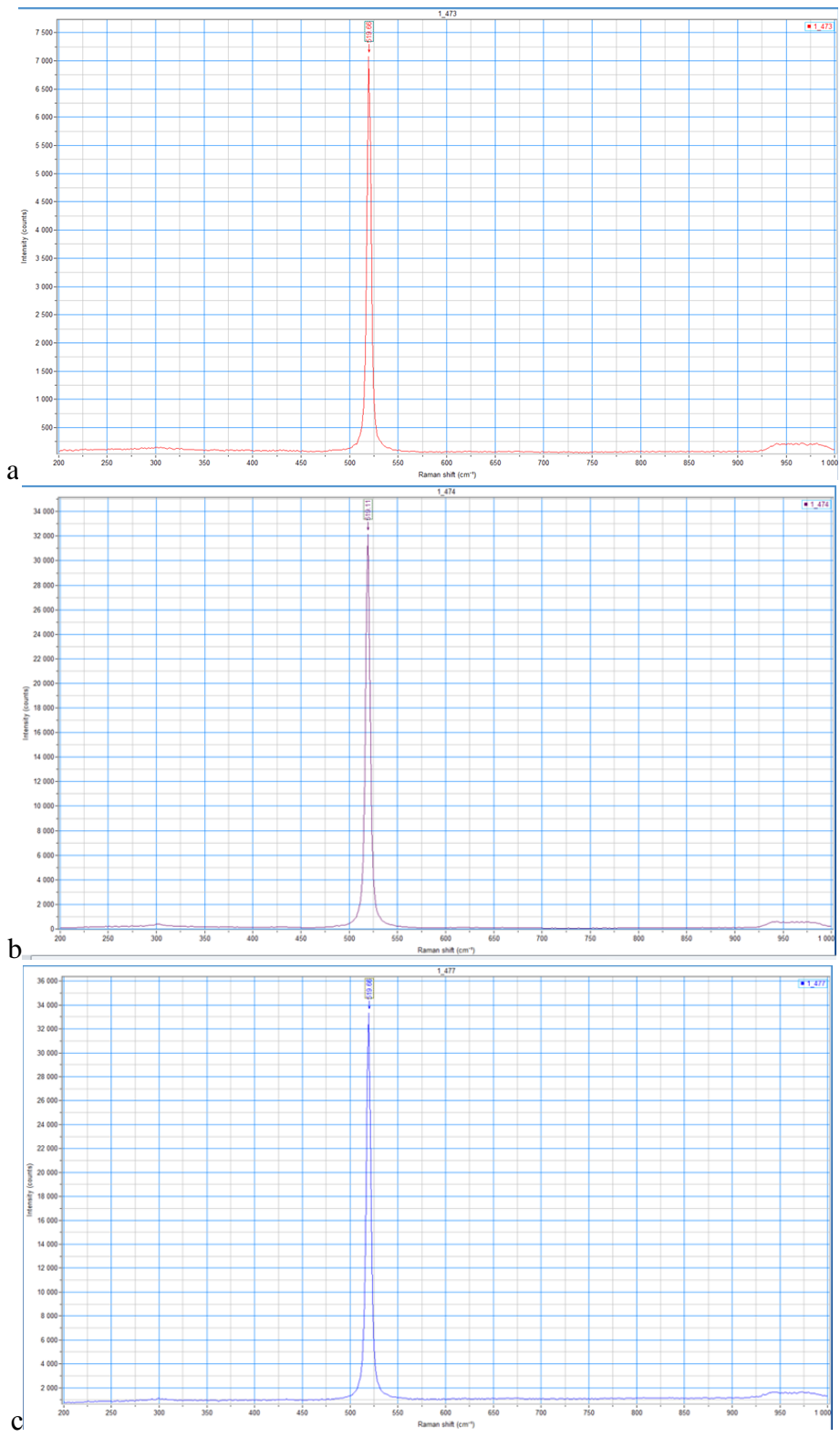

a

b

c

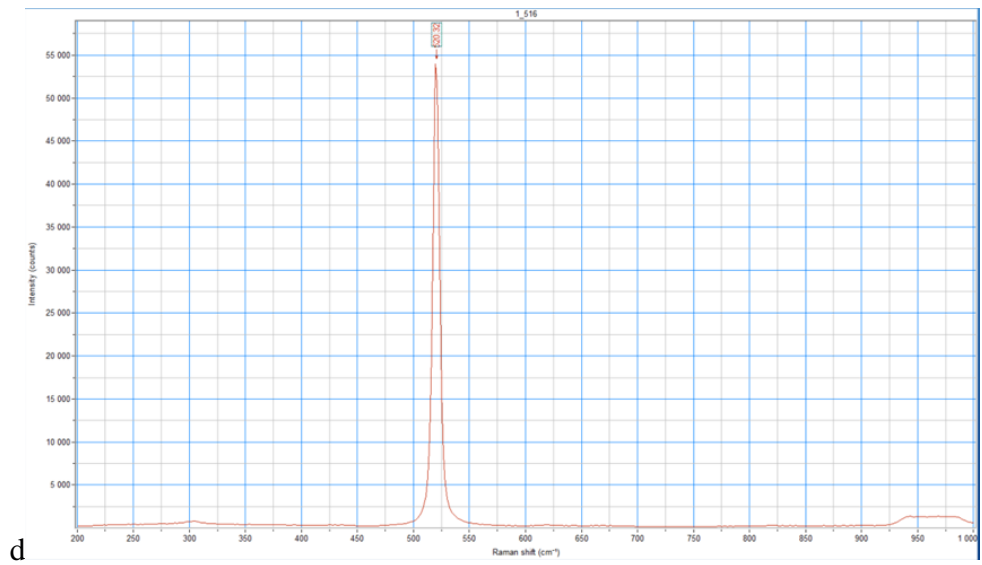

d

Figure 1A. Raman spectra of samples: c-Si –a), PS/AE –b), PS/MACE –c) , PS/MACE+AE –d).

**The EDS- spectra**

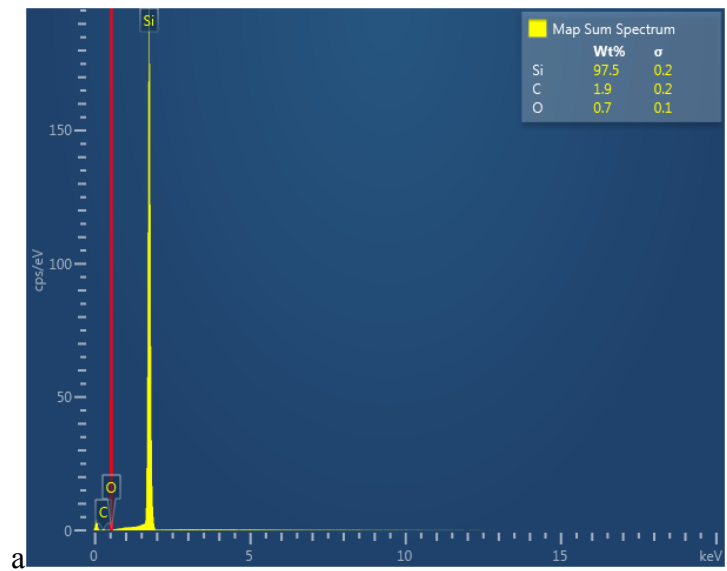

a

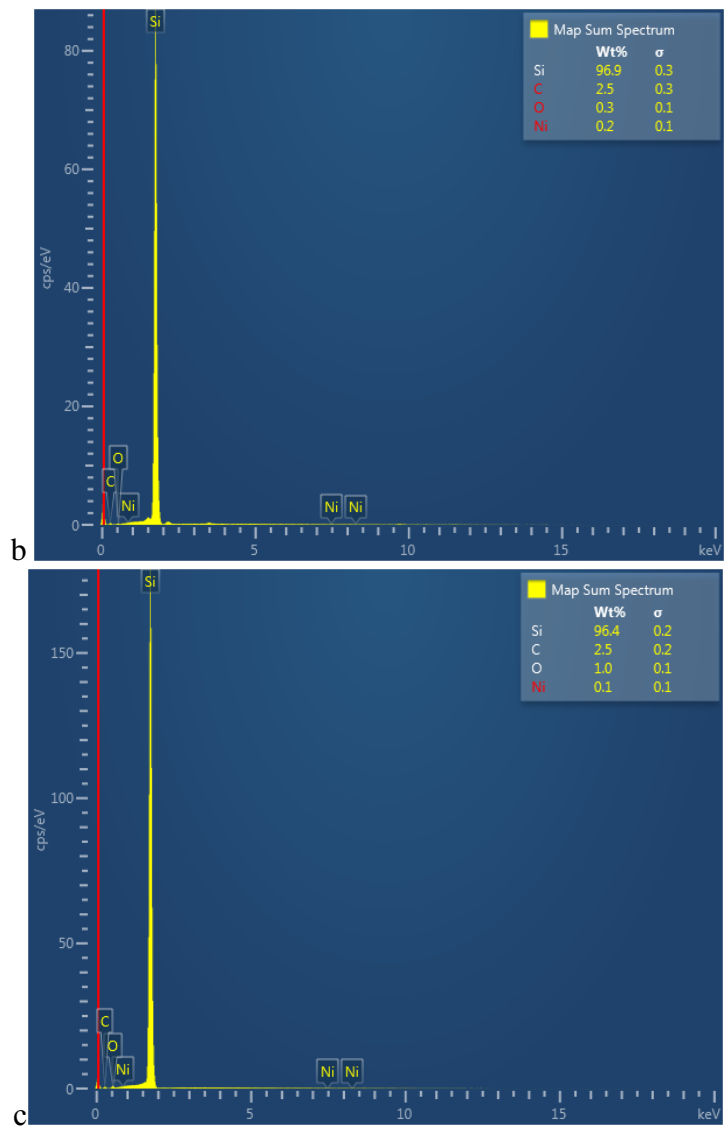

Figure 2A. The EDS- samples spectra: a) PS/AE, b) PS/MACE, c) PS/MACE+AE.

**TGA and DSC spectra**

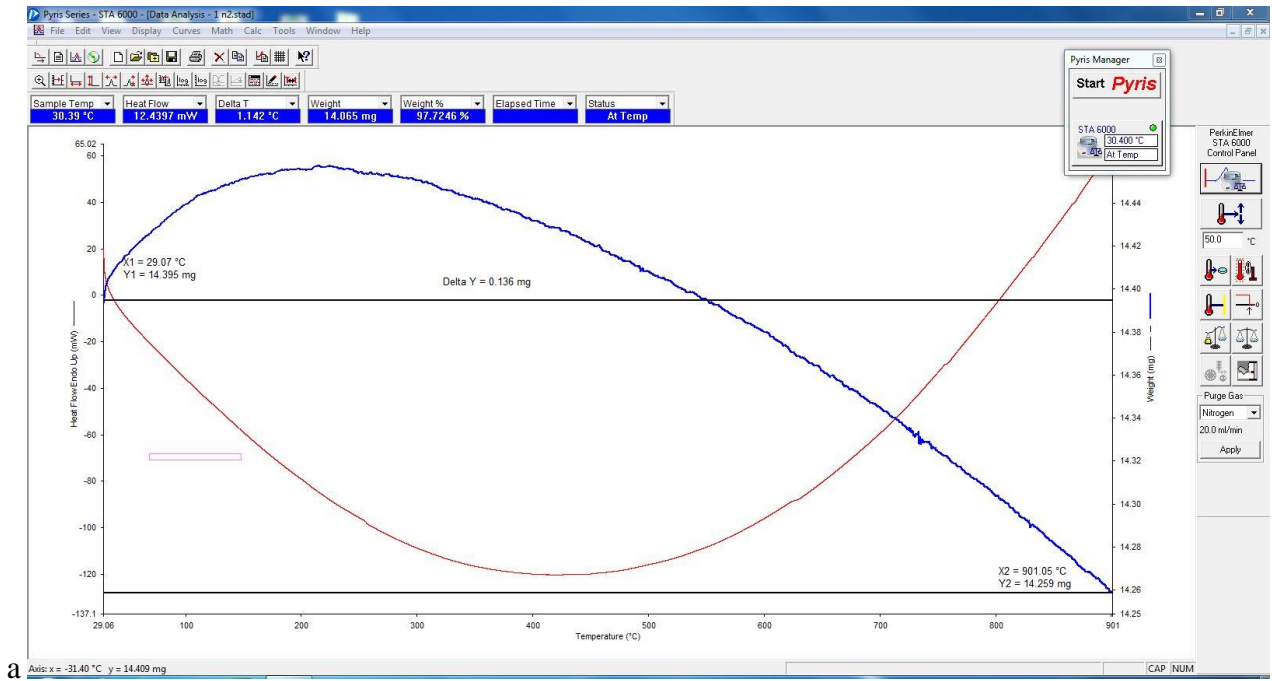

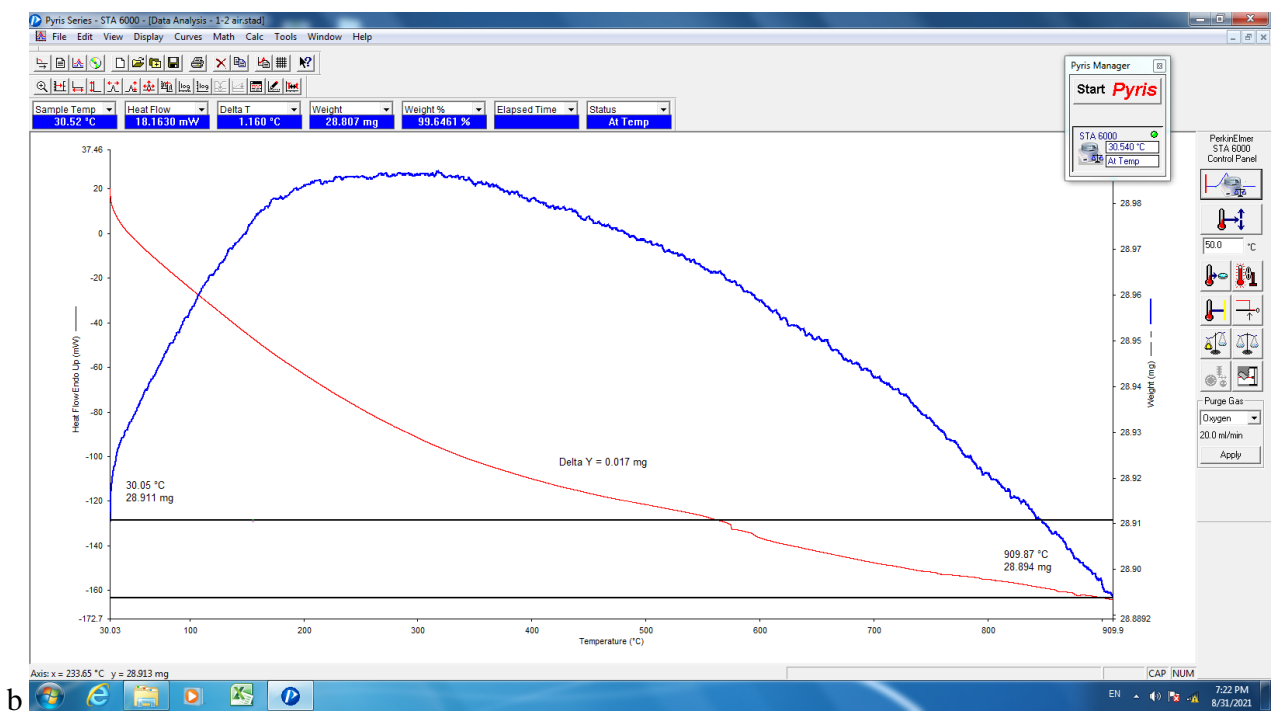

Figure 3A. TGA (blue) and DSC (red line) spectra of the initial c-Si: at nitrogen atmospheres - a), and compressed air - b).

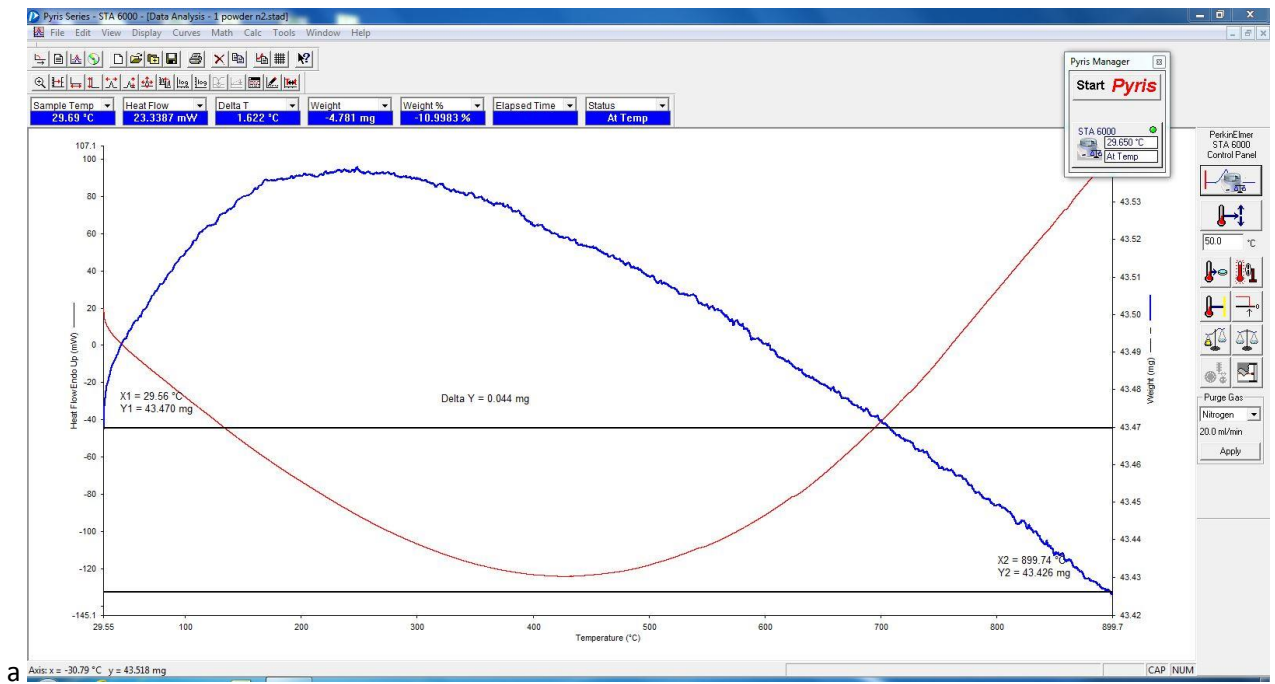

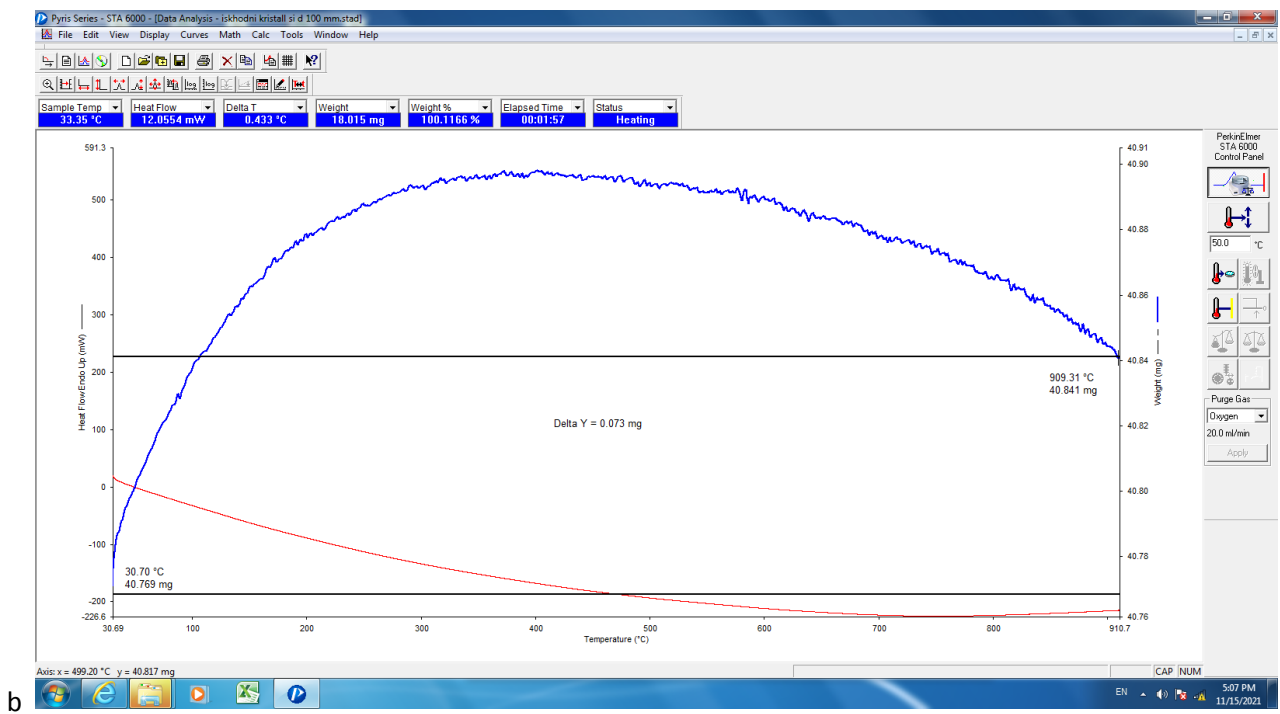

Fig. 4A. TGA (blue) and DSC (red line) spectra of powder-Si: nitrogen – a), air – b).

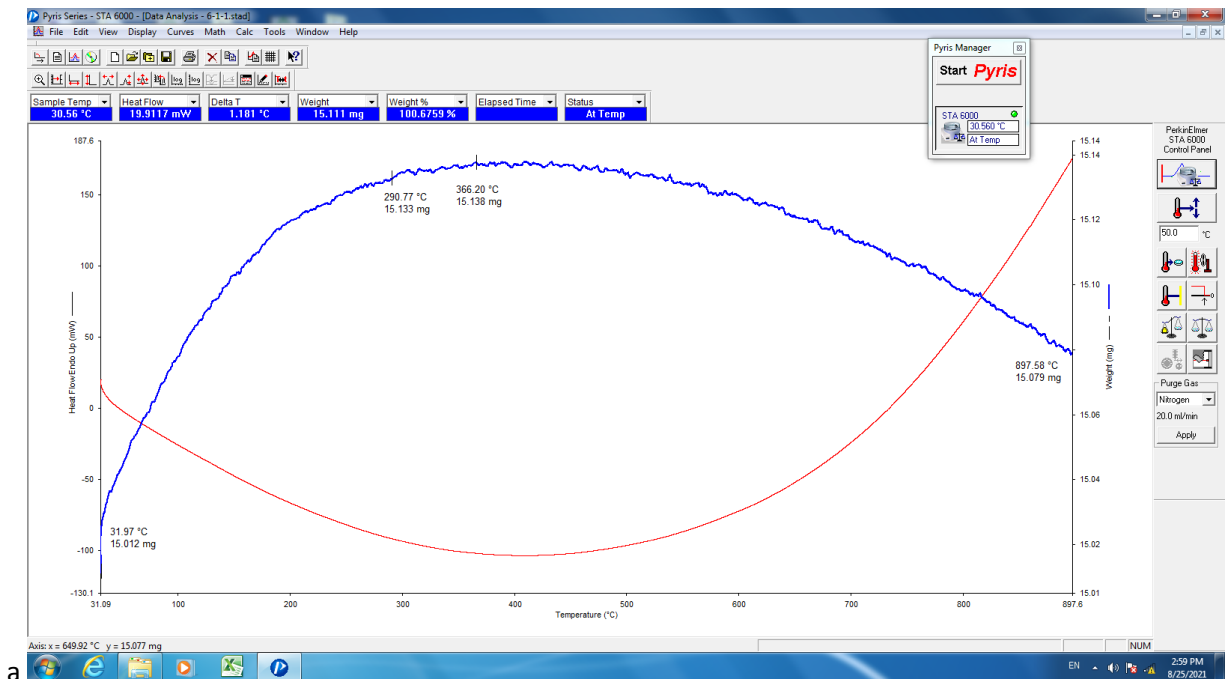

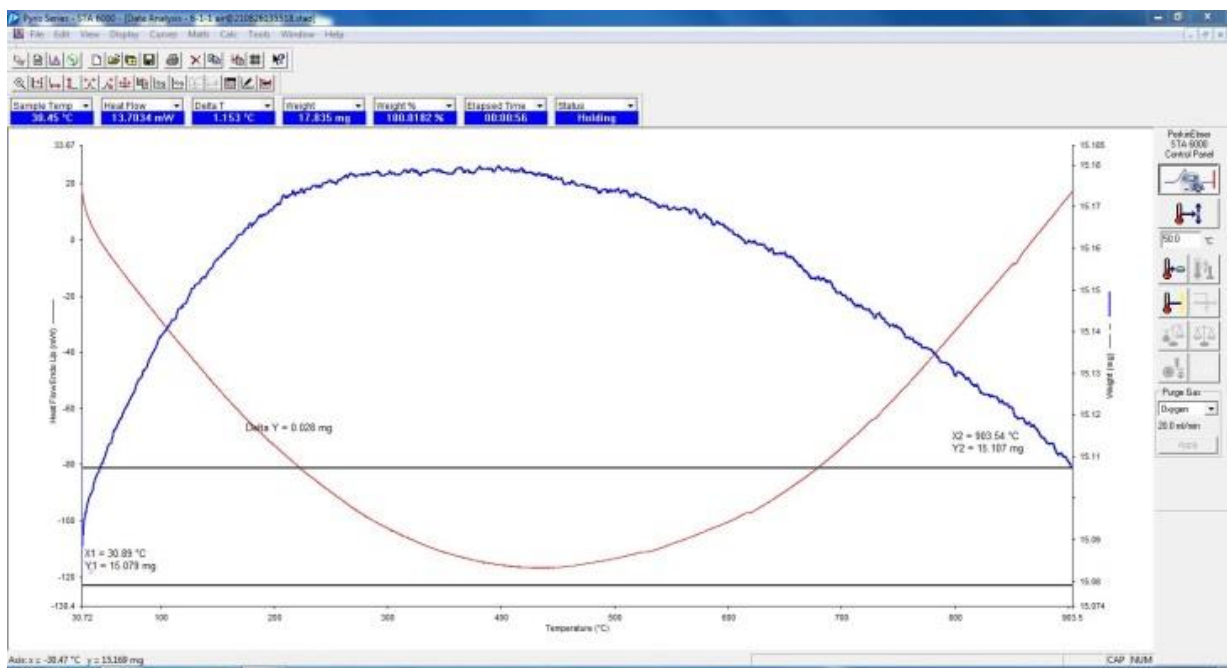

Fig. 5A. TGA (blue) and DSC (red) spectra. Porous silicon PS/AE: nitrogen –a), air – b).

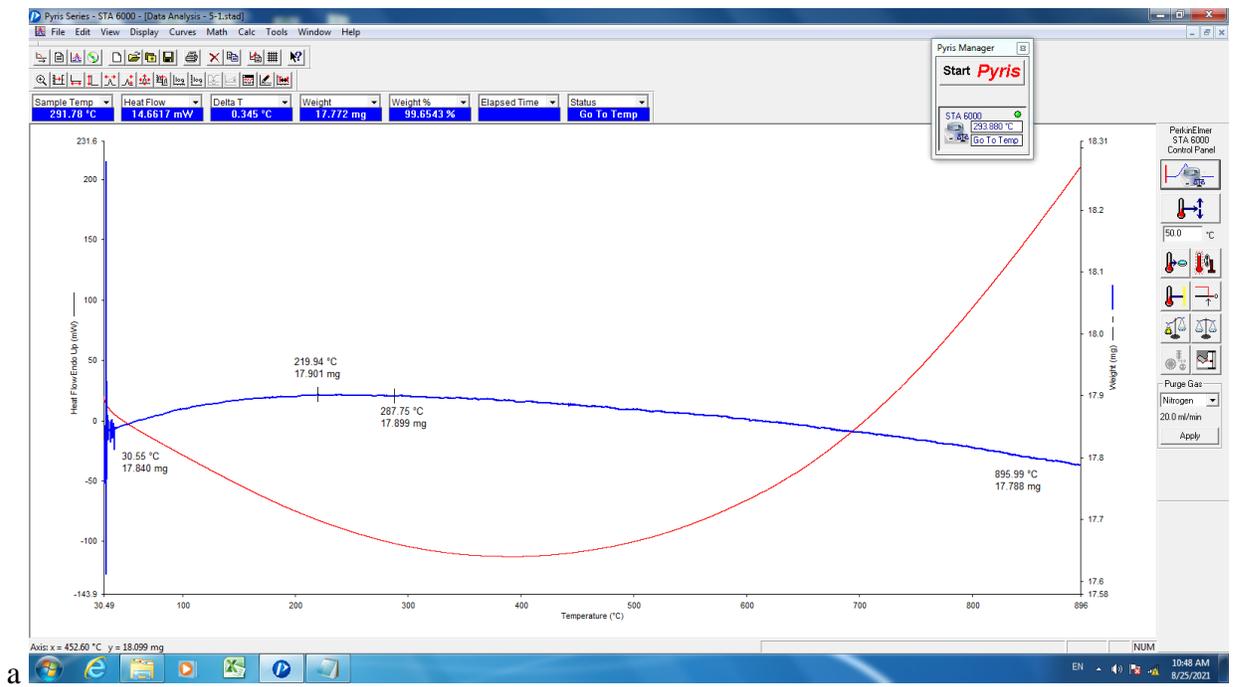

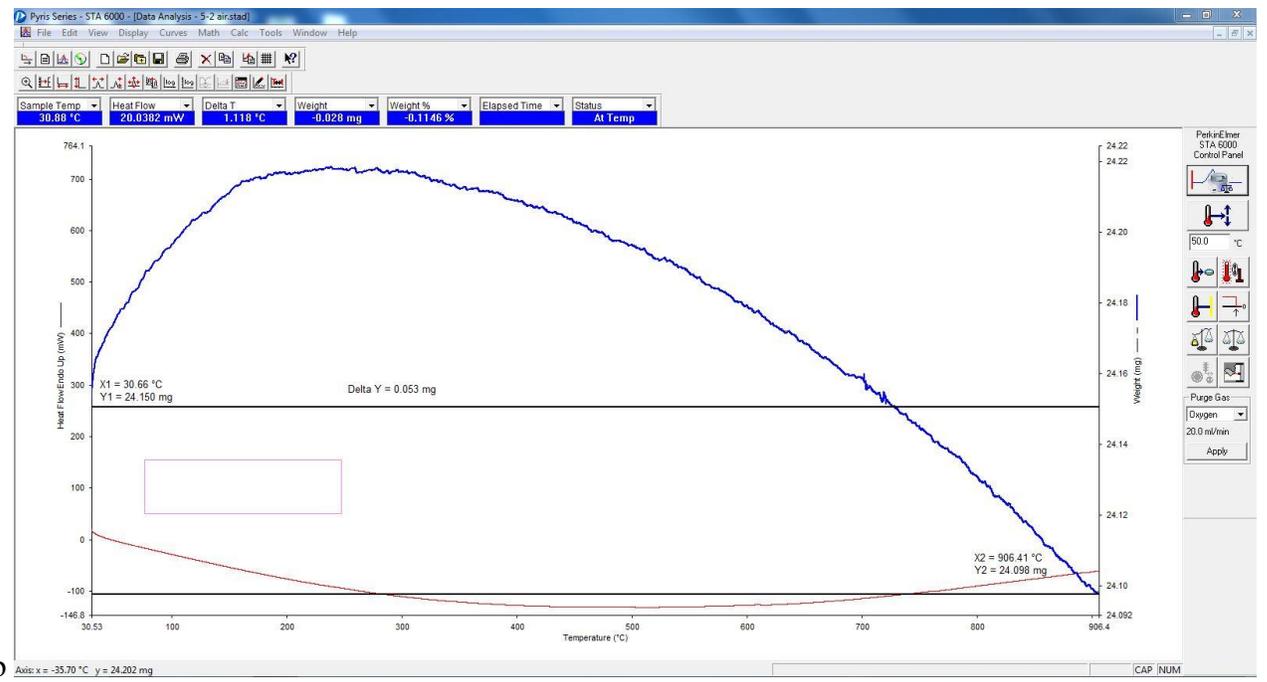

Fig. 6A. TGA (blue) and DSC (red) spectra. Porous silicon PS/MACE: nitrogen – a), air – b).

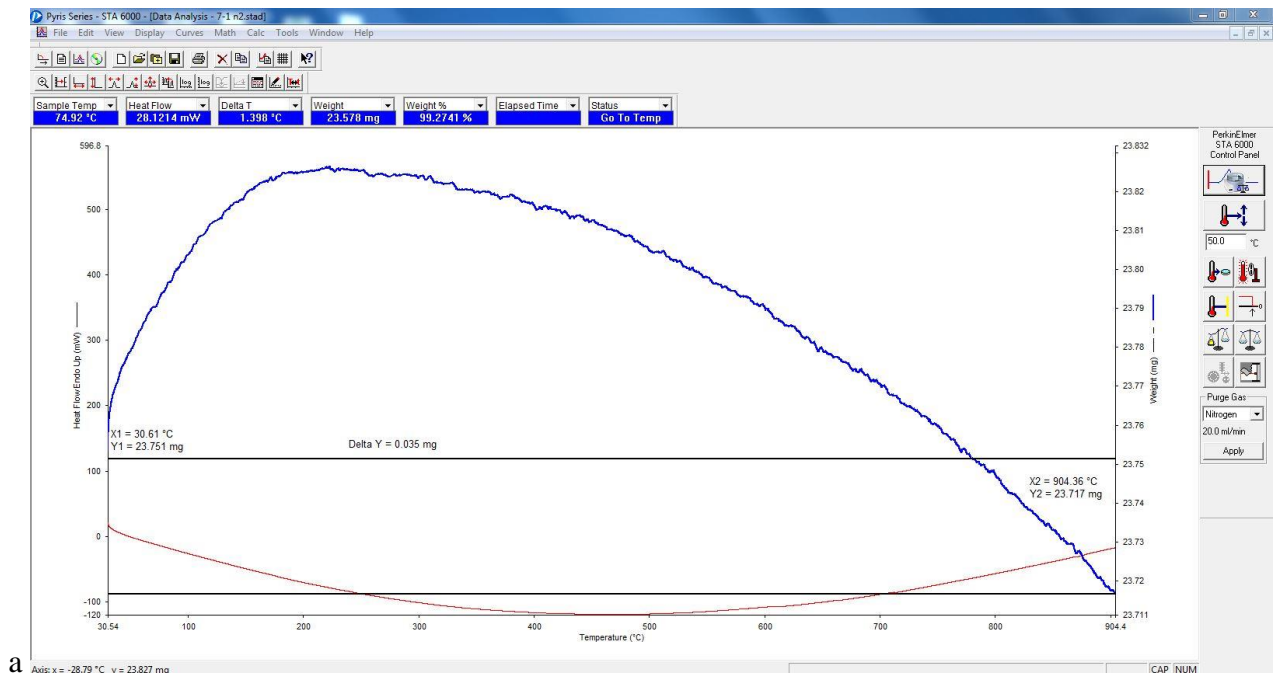

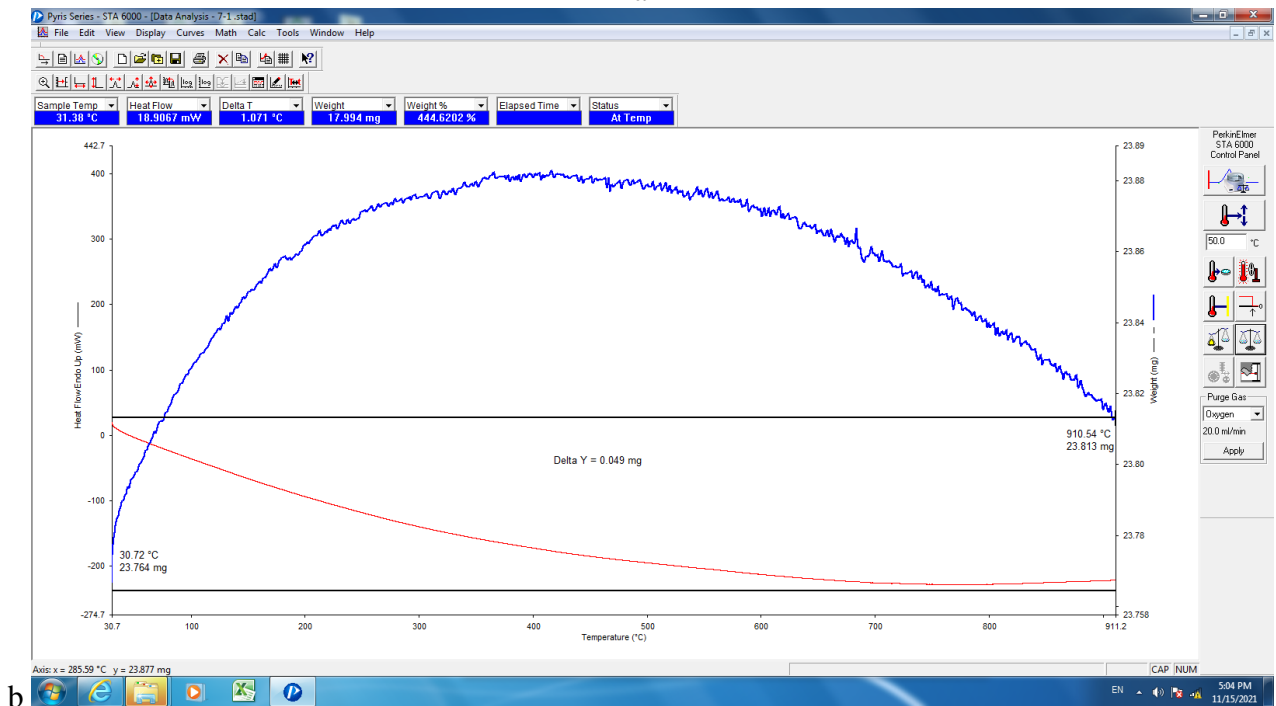

Fig. 7A. TGA (blue) and DSC (red) spectra. Porous silicon PS/MACE+AE: nitrogen –a), air – b).

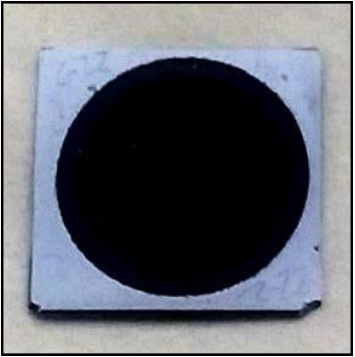

20170712_110405 ПК.mp4

Fig. 8.A. Photo of the PS/EC-sample –a) and video of ethanol splitting after EC- etching- b).